\documentclass[journal,draftclsnofoot,onecolumn,12pt,twoside]{IEEEtran}

\usepackage{ifpdf}
\ifCLASSINFOpdf
  \usepackage[pdftex]{graphicx}
  \DeclareGraphicsExtensions{.pdf,.jpeg,.png}
\else
  \usepackage[dvips]{graphicx}
  \DeclareGraphicsExtensions{.eps}
\fi

\ifCLASSOPTIONcaptionsoff
  \usepackage[nomarkers]{endfloat}
 \let\MYoriglatexcaption\caption
 \renewcommand{\caption}[2][\relax]{\MYoriglatexcaption[#2]{#2}}
\fi

\usepackage{color}
\usepackage{cite}
\usepackage[cmex10]{amsmath}
\usepackage{amsfonts}
\usepackage{mathtools}
\usepackage{amssymb}
\usepackage{amscd}
\usepackage{mathtools}
\mathtoolsset{showonlyrefs}
\usepackage{stackrel}
\usepackage{multirow}
\usepackage{dsfont}
\usepackage{leftidx,tensor}
\usepackage[algo2e]{algorithm2e}
\SetAlgoCaptionSeparator{.}
\SetAlFnt{\small}
\SetAlCapFnt{\small}
\SetAlCapNameFnt{\small}
\usepackage{units}

\newcommand{\abs}[1]{\left|{#1}\right|}

\newcommand{\mbf}[1]{\mathbf{#1}}

\newcommand{\E}[1]{{\mathbb{E}\!\left[{#1}\right]}}

\newcommand{\nth}[1]{{#1}{\text{th}}}

\newcommand{\cpdf}[2]{\mathrm{p}\!\left({#1}\left.\!\right|{#2}\right)}

\renewcommand{\IEEEQED}{\IEEEQEDclosed}

\begin{document}

%
\title{Multi-User MIMO Receivers With Partial State Information}


\author{Ahmad Gomaa, Louay M.A. Jalloul,~\IEEEmembership{Senior~Member,~IEEE,}, Krishna S. Gomadam, Djordje Tujkovic,
        and Mohammad~M.~Mansour,~\IEEEmembership{Senior~Member,~IEEE}}


\maketitle

%
\begin{abstract}
\boldmath
We consider a multi-user multiple-input multiple-output (MU-MIMO) system that uses orthogonal frequency division multiplexing (OFDM). Several receivers are developed for data detection of MU-MIMO transmissions where two users share the same OFDM time and frequency resources. The receivers have partial state information about the MU-MIMO transmission with each receiver having knowledge of the MU-MIMO channel, however the modulation constellation of the co-scheduled user is unknown. We propose a joint maximum likelihood (ML) modulation classification of the co-scheduled user and data detection receiver using the max-log-MAP approximation. It is shown that the decision metric for the modulation classification is an accumulation over a set of tones of Euclidean distance computations that are also used by the max-log-MAP detector for bit log-likelihood ratio (LLR) soft decision generation. An efficient hardware implementation emerges that exploits this commonality between the classification and detection steps and results in sharing of the hardware resources. Comparisons of the link performance of the proposed receiver to several linear receivers is demonstrated through computer simulations. It is shown that the  proposed receiver offers \unit[1.5]{dB} improvement in signal-to-noise ratio (SNR) over the nulling projection receiver at $1\%$ block error rate (BLER) for $64$-QAM with turbo code rate of $1/2$ in the case of zero transmit and receiver antenna correlations. However, in the case of high antenna correlation, the linear receiver approaches suffer significant loss relative to the optimal receiver.
\vspace{-0.05in}

\end{abstract}
\IEEEpeerreviewmaketitle

\vspace{-0.2in}

\section{Introduction}\label{s:intro}\vspace{-0.05in}
As the demand for wireless communication increases, there is a greater challenge for improving network capacity. Allocating more spectrum by the aggregation of several component carriers is a method used to increase the delivered peak data rate and capacity \cite{2010_Yuan}. However, carrier aggregation comes at a substantial capital expenditure as the cellular network operators require first the acquisition of new spectrum and second the associated high network deployment costs to bring these new networks to commercial service. An alternative to spectrum expansion is to exploit the spatial dimension using multiple antennas at the transmitter and receiver, also referred to as MIMO \cite{2003_Paulraj}, which has been successfully used to increase the link throughput and network capacity for several wireless communications standards \cite{802.11,802.16,WCDMA,LTE_36.211}. Amongst the many transmission modes of MIMO \cite{2009_Lee}, multi-user MIMO (MU-MIMO) was proposed as a method for increasing the capacity of a wireless network \cite{2011_Duplicy}. In the LTE standard as an example, which uses MIMO and orthogonal frequency-division multiplexing (OFDM), multiple users on the downlink or uplink can be scheduled for transmission on the same physical resource blocks (PRBs) \cite{2011_Sesia}. The users are paired together such that their cascaded precoder-channel vectors are as orthogonal as possible. Therefore, the capacity is doubled by having the two users share the same time and frequency resource elements. In the downlink, the base station sends pilots over the resource blocks where the users are co-scheduled which are used to estimate the full MU-MIMO channel. For example, in transmission modes 7, 8, and 9 of the LTE standard, the base station transmits pilots with orthogonal cover codes that enable the estimation of the desired and co-scheduled users' channels. The inability of the two users' signals to appear as orthogonal waveforms to the receiver stems from: (1) the finite set of available precoders specified by the standard (to satisfy limited information feedback to be transmitted in the case of frequency division duplex systems) result in non-orthogonality between the users, and (2) during the potential delay between when users are co-scheduled and when the MU-MIMO transmission occurs, the channel state would have changed and thus the orthogonality between the users is lost, typical in the case of high Doppler scenarios. The receivers have only partial state information, since each user receiver has the ability to estimate the MU-MIMO channel, however the modulation constellation of the co-scheduled user remains to be unknown.

Several receiver processing methods at the user equipment (UE) have been proposed for MU-MIMO systems. One approach is to treat the co-scheduled user in MU-MIMO just as additional noise. This type of processing results in sub-optimal receiver performance as depicted in \cite{2011_Duplicy}. Another approach is to employ linear processing using minimum mean-squared error (MMSE) or interference rejection combining (IRC) as shown in \cite{2011_Bai}, where only the knowledge of the channel estimate of the co-scheduled interfering user is used in the detection of the desired user.  In~\cite{2011_Ghaffar_a,2011_Ghaffar_b}, the constellation size of the interfering user's signal is presumed to be 16-QAM regardless of its actual size, and without making any attempt to estimate it, an maximum likelihood (ML) detector is used to separate the two users.

In this work, we consider optimal detection methods for MU-MIMO systems. We argue that detection performance can be enhanced if the UE receiver treats the co-scheduled user's signal as a constrained unknown to be estimated rather than just as additional random noise. We employ joint/conditional ML detection, where the constellation size of the co-scheduled user's signal needs to be estimated first before symbol detection and decoding are performed. We address the co-scheduled user's constellation size estimation problem as part of advanced MU-MIMO detection techniques. We show that the optimal MU-MIMO detector can be \emph{efficiently} implemented with a slight modification of the ML MIMO detector.

The rest of the paper is organized as follows. The system model is described in Section \ref{s:system_model}. Linear receiver methods are discussed in Section~\ref{s:linear_detect}, while the proposed joint optimal constellation size estimation and detection scheme is presented in Section~\ref{s:proposed_scheme}. In Section~\ref{s:implementation}, an efficient hardware architecture for the proposed MU-MIMO receiver is presented. Section~\ref{s:sim} discusses simulation results, while Section~\ref{s:conclusion} concludes the paper.

\textit{Notation}: Unless otherwise stated, lower case and upper case bold letters denote column vectors and matrices, respectively; $\mbf{I}_{N_r}$ denotes the identity matrix of size $N_r$; $(\cdot)^*$ denotes the complex conjugate transpose operation; $(\cdot)^T$ denotes the transpose operation; and $\E{Z}$ is the expectation of the random variable $Z$.

\vspace{-0.1in}
\section{System Model}\label{s:system_model}
Consider a generic OFDM-based MU-MIMO system where $K$ users are scheduled on the same PRBs. Denote by $N$ the number of tones in each PRB, and assume that each UE is equipped with $N_r$ antennas. The received frequency-domain complex signal $\mbf{y}_i\!\in\!\mathcal{C}^{N_r\times 1}$ at the UE of interest on the $\nth{i}$ resource element over which the $K$ users are scheduled is given by
\begin{equation}\label{eq:sys_model}
  \mbf{y}_i = \mbf{H}_i \mbf{x}_i + \mbf{n}_i,~i\!=\!1,2,\cdots,N,
\end{equation}\\[-2.1em]
where $\mbf{H}_i \!\in\!\mathcal{C}^{N_r \times K}$ is the complex channel matrix, $\mbf{x}_i$ denotes the transmitted $K\!\times\! 1$ QAM symbol vector for $K$ users, and $\mbf{n}_i\!\in\!\mathcal{C}^{N_r\times 1}$ denotes thermal noise modeled as a zero-mean complex Gaussian random vector with covariance matrix $\E{\mbf{n}_i\mbf{n}^*_i}=\mbf{R}=\sigma^2 \mbf{I}_{N_r}$.

We consider the practical example where $K\!=\!N_{r}\!=\!2$, so that the received signal at the $\nth{i}$ resource element can be written as follows\vspace{-0.04in}
\begin{equation}\label{eq:yi_expanded}
    \mbf{y}_i = \mbf{h}_1^{(i)}x_{1}^{(i)} + \mbf{h}_2^{(i)} x_{2}^{(i)} + \mbf{n}_i,~i=1,\cdots,N,
\end{equation}\\[-1.75em]
where $\mbf{H}_i \!\triangleq\! \bigl[ \mbf{h}_1^{(i)} ~\mbf{h}_2^{(i)} \bigr]$, and $\mbf{h}_1^{(i)}$ and $\mbf{h}_2^{(i)}$ denote the cascade of the channel and the precoders of the user 1 and user 2, respectively.

Let user 1 denote the user of interest while user 2 denote the interfering (co-scheduled) user whose constellation size is unknown to user 1. Denote $M_{S}$ and $M_{I}$ as the constellations of user 1 and user 2, respectively, and $N$ as the number of resource elements over which $M_{I}$ is constant. The symbols $x_{1}^{(i)}$ and $x_{2}^{(i)}$ represent $\text{log}_2(M_S)$ and $\text{log}_2(M_I)$ coded bits from a channel encoder (e.g., turbo code), respectively. The $\nth{j}$ bit of symbol $x_{1}^{(i)}$ is denoted by $b_{ij}$.

\vspace{-0.15in}
\section{Linear Receiver Methods}\label{s:linear_detect}
Two linear detection methods for MU-MIMO based on IRC are considered. They differ in terms of their complexity and performance.\vspace{-0.2in}

\subsection{Covariance-based Linear Receiver} \label{s:linear_cov}
This receiver structure estimates the interference covariance matrix using knowledge of only the desired user's channel, thus having least knowledge about the co-scheduled user. Denote $x_{1}^{(p)}$ and $\mbf{y}_p$ as the known pilot and received vector, respectively, on the $\nth{p}$ tone, where $p\!=\!1,2,\cdots,N_p$ and $N_p$ is the number of pilot tones amongst the set of $N$ tones over which the two users are co-scheduled. Then, the interference sample covariance matrix is estimated as\vspace{-0.075in}
\begin{equation}
  \mbf{R_{uu}} = \frac{1}{N_p} \sum_{p=1}^{N_p} \mbf{u}_p \mbf{u}_p^*,
\end{equation}\\[-1.1em]
where $\mbf{u}_p = \mbf{y}_p - \mbf{h}_1^{(p)}x_{1}^{(p)}$. Note that there is a single interference covariance matrix generated per PRB, i.e. $\mbf{R_{uu}}$ is a sample covariance, obtained by averaging over the pilot tones in the PRB and thus is independent of the tone index $i$. The weight vector for the linear receiver is given by \cite{2003_Paulraj}\vspace{-0.075in}
\begin{equation} \label{eq:cov_linear}
    \mbf{w}_{1,\mathsf{cov}}^{(i)} = \mbf{R}^{-1}_{\mbf{uu}} \mbf{h}_1^{(i)}.
\end{equation}
The desired user receiver applies this weight vector \eqref{eq:cov_linear} to the received vector to obtain $\mbf{w}_{1,\mathsf{cov}}^{(i)*} \mbf{y}_i\!\triangleq z_{1,\mathsf{cov}}$, which is a scalar used to generate the soft decision metric for decoding.

\vspace{-0.2in}
\subsection{Linear IRC Receiver} \label{s:linear_mmse}
An improvement over the covariance-based linear receiver is obtained wherein the UE receiver uses estimates of \emph{both} the desired channel as well as the co-scheduled user's channel. We arrive at an IRC weight vector that is given by \cite{2011_Bai}
\begin{equation}
    \mbf{w}_{1,\mathsf{IRC}}^{(i)} = \left( \mbf{h}_2^{(i)} \mbf{h}_2^{(i)*} + \mbf{R}_i \right)^{-1} \mbf{h}_1^{(i)}.
\end{equation}\\[-1.5em]
The desired user receiver processing is obtained by
\begin{equation}
z^{(i)}_{1,\mathsf{IRC}} = \mbf{w}_{1,\mathsf{IRC}}^{(i)*} \mbf{y}_i,
\end{equation}\\[-1.5em]
which can be approximated by a complex Gaussian random variance, i.e.
$\cpdf{\!z^{(i)}_{1,\mathsf{IRC}}}{x^{(i)}_1,\mbf{h}^{(i)}_1,\mbf{h}^{(i)}_2} \sim \mathcal{CN} \left( \mu_{z},\nu_{z}^2 \right)$ where the mean and variance are given by
\begin{align}
\mu_{z} &= \mbf{w}_{1,\mathsf{IRC}}^{(i)*} \mbf{h}^{(i)}_1 x^{(i)}_1\\[-0.25em]
\nu_{z}^2 &= \mbf{w}_{1,\mathsf{IRC}}^{(i)*} \left( \mbf{h}^{(i)}_2 \mbf{h}^{(i)*}_2 + \mbf{R}_i \right) \mbf{w}_{1,\mathsf{IRC}}^{(i)}.
\end{align}\\[-1.5em]
The LLR of the $\nth{j}$ bit of the desired user QAM symbol $x^{(i)}_1$ using the max-log-MAP approximation is given by
\begin{equation}\label{eq:LLR_IRC}
\Lambda \left( b_{ij} \right) \!=\!
 \frac{1}{\nu_{z}^2}\! \left[\!
 \underset{x_{1}^{(i)}\in M^{(-1)}_S}{\mathop{\min }}\!
 \left| z^{(i)}_{1,\mathsf{IRC}} - \mu_{z} \right|^2
  \!-\!  \underset{x_{1}^{(i)}\in M^{(+1)}_S}{\mathop{\min }}\! \left| z^{(i)}_{1,\mathsf{IRC}} - \mu_{z} \right|^2 \!\right]
\end{equation}
where
$M^{(+1)}_S \!=\!\left\{x_{1}^{(i)}:b_{ij}\!=\!+1 \right\}$ and $M^{(-1)}_S \!=\!\left\{x_{1}^{(i)}:b_{ij}\!=\!-1 \right\}$.

\section{Reduced Complexity Constellation Size Estimation}\label{s:linear_const_det}
We estimate the constellation size of the co-scheduled user by nulling out the \emph{unknown} signal of the user of interest using a linear filter. The equivalent system after applying this linear filter to the received vector is given by the following single-input single output relation
\begin{equation}\label{eq:yi_tild}
    \tilde{y}_{i} = \mbf{g}_{i}^{*} \mbf{y}_{i}^{} = \mbf{g}_{i}^{*} \mbf{h}_{2}^{(i)} x_{2}^{(i)} + \mbf{g}_{i}^{*} \mbf{n}_{i}^{} \triangleq {a_i^{}}\text{ }x_{2}^{(i)}+{\tilde{n}_i},
\end{equation}\\[-1.75em]
where $\mbf{g}_{i}\!\in\!\mathcal{C}^{2\times 1}$ is chosen to be orthogonal to $\mbf{h}_{1}^{(i)}$, i.e., $\mbf{g}_{i}^{*}\mbf{h}_{1}^{(i)}=0$. The nulling filter is given by
$\mbf{g}_{i} = \mbf{G} \mbf{c}$, where $\mbf{c}$ is a $2 \times 1$ combining vector and $\mbf{G}$ is the left null-space projection matrix given by~\cite{1996_Golub}
\begin{equation}\label{eq:proj_matrix_G}
    \mbf{G} = \left( \mbf{I}_2 - \frac{\mbf{h}_{1}^{(i)}\mbf{h}_{1}^{(i)*}}{\|\mbf{h}_{1}^{(i)}\|^2} \right).
\end{equation}
Since $\mbf{G}$ has rank 1, the combining vector can simply be $\mbf{c}=[1~~0]^T$.

Next, the ML estimate of the constellation is obtained based on the filtered received signal $\tilde{y}_{i}$ and can be approximated as
\begin{equation}\label{eq:Null_est_max_log}
    \hat{M}_I \!\approx\! \underset{M_I\in \mathcal{M}}{\mathop{\arg\min }}\left(\! N\log\left( \abs{M_I} \right) + \sum\limits_{i=1}^{N}   \frac{{\underset{x_{2}^{(i)}\in M_I}{\mathop{\min }}}\left| \tilde{y}_{i} - {a}_{i}x_{2}^{(i)} \right|^2}{\mbf{g}^*_i \mbf{R} \mbf{g}_i} \right),
\end{equation}
where $\log(\cdot)$ is the natural logarithmic function and $\mathcal{M}\!\triangleq\!\left\{ \varnothing,\text{4-QAM},\text{16-QAM},\text{64-QAM} \right\}$ denotes the set of allowable constellations that the interferer can assume, including the case when the co-scheduled user is not present, i.e. $\varnothing$. The derivation of \eqref{eq:Null_est_max_log} is omitted here as it follows closely the development of the next section. Once the co-scheduled user constellation, $\hat{M}_I$, is estimated, then the LLR are generated as shown in the next section.

\vspace{-0.1in}
\section{Joint Constellation Size Estimation and Detection}\label{s:proposed_scheme} \vspace{-0.05in}
The key idea is to better exploit the diversity offered by the two receive antennas. We develop an optimal receiver for user 1 based on a step of estimating the modulation constellation of user 2 using the received signal $\mbf{y}_i$ and knowledge of the channel. We derive the ML estimate of the constellation size based on the received signal itself $\mbf{y}_{i}$ instead of a reduced-dimension version obtained as was done in Section~\ref{s:linear_const_det}. The optimal ML estimator will inherently average out the unknown desired signal using knowledge of its constellation size, and yields better results than just nulling out the unknown signal. The new ML estimate of the constellation of the interfering user based on $\left\{\mbf{y}_{i}\right\}_{i=1}^{N}$ is given by
\begin{align}\label{eq:ML_estimate}
    \hat{M}_I =  \underset{M_I\in \mathcal{M}}{\mathop{\arg\max }}\,\mathrm{p}\left( \left. \left\{\mbf{y}_{i}\right\}_{i=1}^{N} \right|\left\{ \mbf{H}_i \right\}_{i=1}^{N},M_S,M_I, \right).
\end{align}
Since $x_{1}^{(i)}, x_{2}^{(i)}$ and $\mbf{n}_{i}$ are independent for  $i\!=\!1,\cdots,N$, the ML estimate of the interferer's constellation can then be written as\small
\begin{align}
  {\hat{M}}_{I}
    &\!=\!  \underset{M_I\in \mathcal{M}}{\mathop{\arg\max}}\prod\limits_{i=1}^{N}{\mathrm{p}\left( \left.
        \mbf{y}_{i} \right|\mbf{H}_i,M_S,M_I \right)} \label{eq:ML_estimate_step1}\\[-0.30em]
    &\!=\! \underset{M_I\in\mathcal{M}} {\mathop{\arg\max}}\!\prod\limits_{i=1}^{N}{\!\!\!\sum\limits_{x_{1}^{(i)}\!\in\! M_S}{\!\!\sum\limits_{x_{2}^{(i)}\!\in\! M_I}{\!\!\!\!\!\mathrm{p}\!\left(\! \left. \mbf{y}_{i} \right|\!\mbf{H}_i\!,M_S\!,M_I\!,x_{1}^{(i)}\!,x_{2}^{(i)} \!\right)\!\mathrm{p}\!\left(\!x_{1}^{(i)}\!\right)\!\mathrm{p}\!\left(\!x_{2}^{(i)}\!\right)}}} \label{eq:ML_estimate_step2}\\[-1.45em]
    &\!=\!\underset{M_I\in \mathcal{M}}{\mathop{\arg\max }}\,\frac{1}{{{\abs{M_{I}}}^{N}}}\prod\limits_{i=1}^{N}{\sum\limits_{x_{1}^{(i)}\in M_S}{\sum\limits_{x_{2}^{(i)}\in M_I}{\!\!\!\!\mathrm{p}\!\left(\! \left. \mbf{y}_{i} \right|\mbf{H}_i,M_S,M_I,x_{1}^{(i)},x_{2}^{(i)} \!\right)}}}\label{eq:ML_estimate_step3}
\end{align}\\[-2em]\normalsize
where $\abs{M_{I}}$ denotes the size of the interfering user constellation. The last equality follows from the assumptions that
\begin{align}\label{eq:prior_prob}
    \mathrm{p}\left(x_{1}^{(i)}\right) = \frac{1}{M_S},\text{ }\text{      and     }\mathrm{p}\left(x_{2}^{(i)}\right) = \frac{1}{M_I},\text{ }\forall i.
\end{align}
Let $\mbf{x}_i\!=\!\bigl[ x_{1}^{(i)}\, x_{2}^{(i)} \bigr]^T$ and $d(\mbf{x}_i)\!=\!\left( \mbf{y}_i\!-\!\mbf{H}_i\mbf{x}_i \right)^*\! \mbf{R}^{-1}\! \left( \mbf{y}_i\!-\! \mbf{H}_i \mbf{x}_i \right)$, we can then write $\hat{M}_I$ as\vspace{-0.1in}
\begin{align}\label{eq:ML_estimate_exp}
    \hat{M}_I = \underset{ M_I \in \mathcal{M}}{\mathop{\arg\max }}\,\frac{1}{\abs{M_I}^{N}}\prod\limits_{i=1}^{N}{\sum\limits_{x_{1}^{(i)}\in M_S}{\sum\limits_{x_{2}^{(i)}\in M_I}{\exp{\left(-d(\mbf{x}_i)\right)}}}}.
\end{align}\\[-1em]
Using the max-log-MAP approximation, we have that
\begin{align}\label{eq:ML_estimate_max_log}
    \hat{M}_I \!\approx\! \underset{M_I\in \mathcal{M}}{\mathop{\arg\min }}\left(\! N\log\left( \abs{M_I} \right) + \sum\limits_{i=1}^{N} {\underset{x_{1}^{(i)}\in M_S,x_{2}^{(i)}\in M_I}{\mathop{\min }}} \!\!\! d(\mbf{x}_i)  \right).
\end{align}
The simplifications to~\eqref{eq:ML_estimate_max_log} will become evident in the receiver implementation Section~\ref{s:implementation}. As seen from \eqref{eq:ML_estimate_max_log}, the modulation classification metric is sum over the set of tones, $N$, over which the co-scheduled user is stationary, of the Euclidean distance between the received vector and the symbol 2-tuple $\bigl[ x_{1}^{(i)}\, x_{2}^{(i)} \bigr]^T$.

Once the co-scheduled user's constellation, $\hat{M}_I$, is estimated, then the LLR of the $\nth{j}$ bit of the desired user QAM symbol $x^{(i)}_1$ using the max-log-MAP approximation is given by \cite{2006_Fitz}
\begin{equation}\label{eq:final_LLR}
\Lambda \left( b_{ij} \right) \!=\!
\underset{x_{1}^{(i)}\in M^{(-1)}_S,x_{2}^{(i)}\in \hat{M}_I}{\mathop{\min }}
d(\mbf{x}_i) - \!\!\underset{x_{1}^{(i)}\in M^{(+1)}_S,x_{2}^{(i)}\in \hat{M}_I}{\mathop{\min }} d(\mbf{x}_i),
\end{equation}
where $M^{(+1)}_S$ and $M^{(-1)}_S$ are defined in Section \ref{s:linear_mmse}. As seen from \eqref{eq:final_LLR}, computing the LLRs involves the same Euclidean distance computations as those needed for the co-scheduler user's constellation estimation in \eqref{eq:ML_estimate_max_log}. This fact will be exploited in the hardware implementation in the joint constellation classification and data detection.

\section{MU-MIMO Receiver Hardware Implementation}\label{s:implementation} 
\begin{figure}[t]
\centering
\includegraphics[scale=1.3]{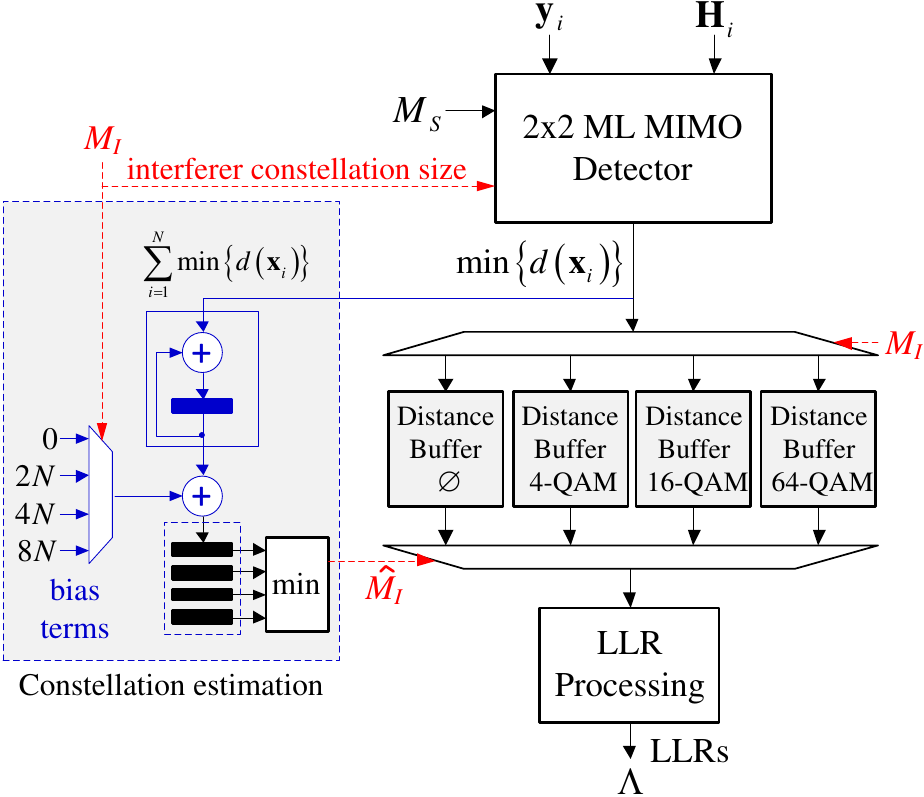}
\caption{Architecture for a $2\times 2$ MU-MIMO detector}
\label{f:architecture}
\end{figure}

\vspace{-0.05in}
Figure~\ref{f:architecture} shows an optimized architecture for a $2\!\times\! 2$ MU-MIMO detector, constructed using an ML MIMO detector as its core. The MIMO detector detects the received signal $\mbf{y}_i$ assuming all 4 possible choices of the interferer's constellation. It generates 4 corresponding lists of minimum distance metrics $d(\mbf{x}_i)$ and their associated symbol vectors $\mbf{x}_i$ for all the $\abs{M_I}$ possible hypothesis of interfering user constellation, with $x_1^{(i)}\!\in\!M_S$. These distances and symbols are stored in four buffers each of size $\abs{M_S}$ as shown in the figure.

For each tone to be detected, the minimum distance from each list is passed to an adder that accumulates the minimum distances over a span of $N$ tones, during which the interferer modulation is assumed to be static. The resulting 4 minimum accumulated distances for each interferer hypothesis are stored in a buffer. The minimum from this buffer is used to identify the interferer's constellation, and the corresponding stored distances in the buffers are selected and forwarded for LLR processing according to~\eqref{eq:final_LLR}.

Note that since the interferer's modulation constellation remains static over $N$ tones for a duration of 1 subframe in LTE (14 OFDM symbols), the particular choice of $N\!=\!12$ results in substantial savings in computations. The detector only needs to run in the above mode to identify the interferer's constellation for one OFDM symbol in the subframe. It can then switch to back to normal ML detection mode (without modulation classification) to generate the LLRs for the remaining 13 OFDM symbols for the user of interest $x_1^{(i)}$.

Taking the LTE scenario for hardware complexity analysis, the total number of possible tones in 1 PRB in a subframe is $12\!\times\!14\!=\!168$. Of these tones, 28 are reserved for pilots (for cell specific reference signals and for UE specific pilots to support the MU-MIMO transmission mode), and 140 for data. In the hardware architecture of Fig. \ref{f:architecture}, the total number of distance computations needed to generate the LLRs from these 140 data tones is $140\!\times\!\abs{M_S}$. This corresponds to an increase of only $22.8\%$ compared to the distances computed by an ML detector with perfect knowledge of the interferer.

\section{Simulation Results}\label{s:sim}
\begin{figure}
\centering
\includegraphics[scale=1]{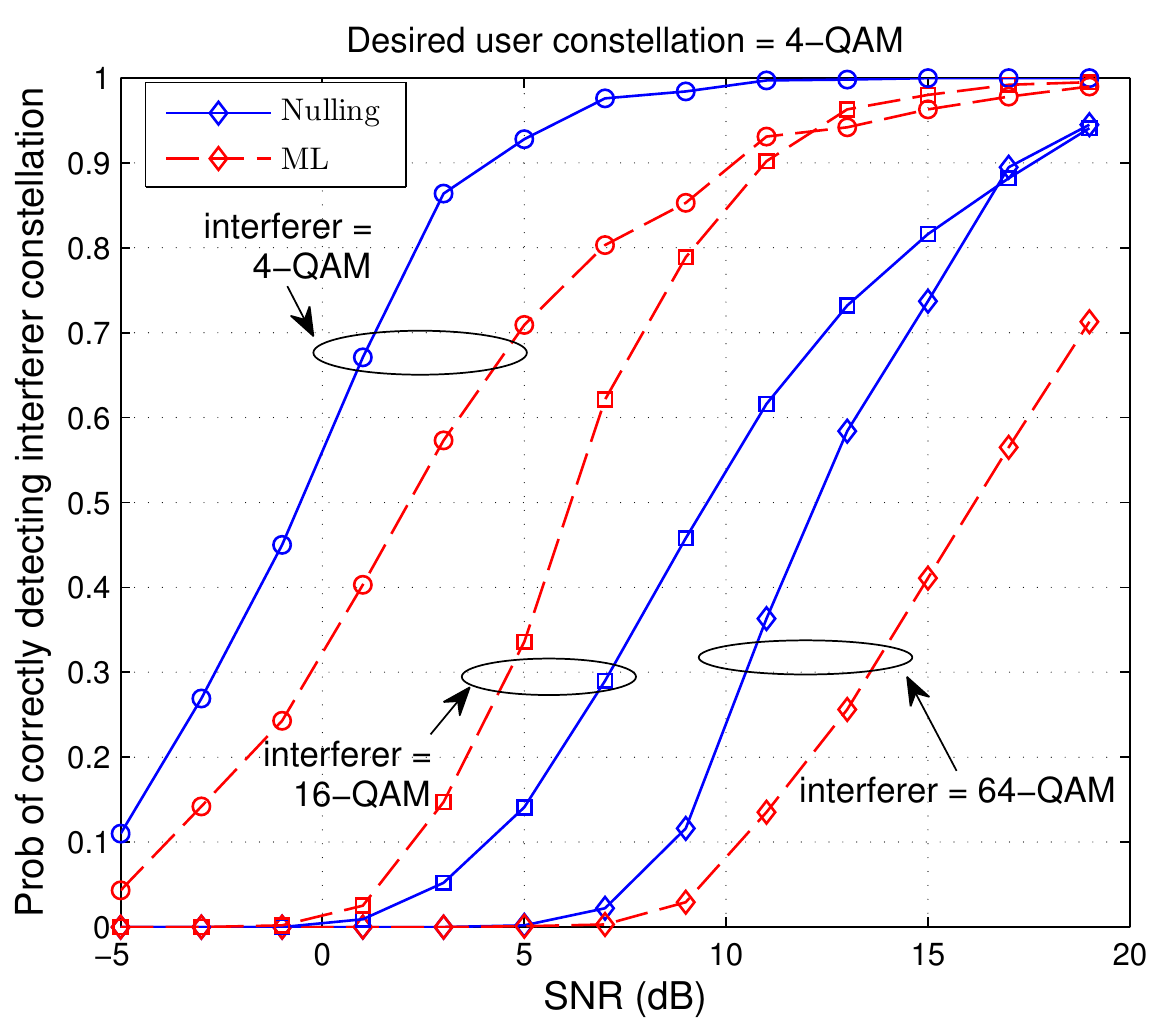}
\caption{Probability of correct interferer modulation constellation detection versus SNR (dB). Solid lines are for nulling approach, dashed are for the ML approach. Desired user constellation is fixed to $4$-QAM and the co-scheduled user constellation is $4,16, 64$-QAM. The channel is i.i.d. block fading.}
\label{f:fading_scenario_sig_4QAM_diffInterfConst}
\end{figure}

\begin{figure}
\centering
\includegraphics[scale=1]{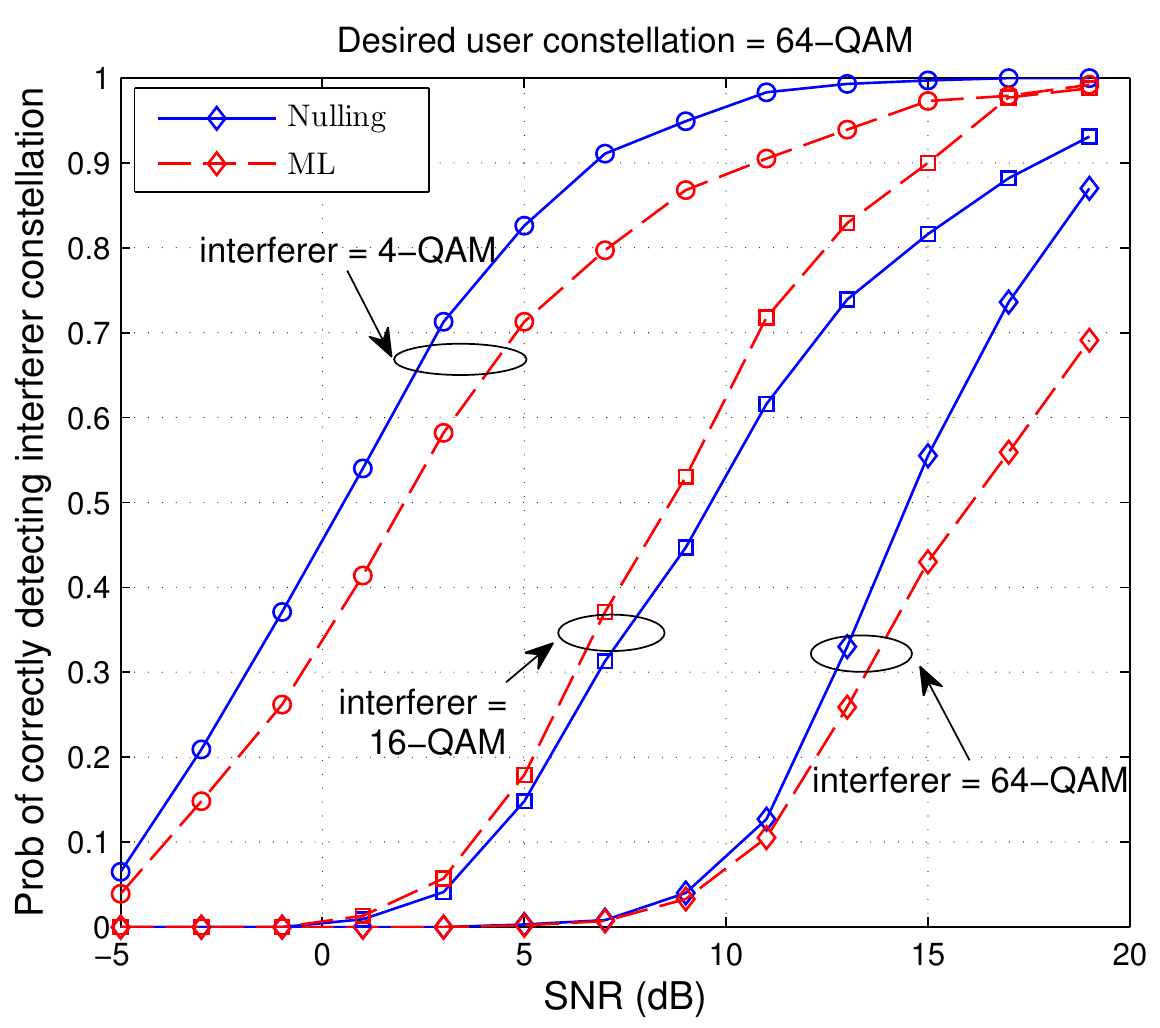}
\caption{Probability of correct interferer modulation constellation detection versus SNR (dB). Solid lines are for nulling approach, dashed are for the ML approach. Desired user constellation is fixed to $64$-QAM and the co-scheduled user constellation is $4,16, 64$-QAM. The channel is i.i.d. block fading.}
\label{f:fading_scenario_sig_64QAM_diffInterfConst}
\end{figure}

The performance of the null projection method of Section \ref{s:linear_const_det} is compared to the proposed method of Section \ref{s:proposed_scheme}. We simulate the probability of correctly detecting the constellation size of the interfering user for different desired and interfering constellation sizes. The received desired user power is assumed to be equal to the received interfering user's power, i.e. the signal-to-interference ratio (SIR) is \unit[0]{dB}. The channel is generated as a zero-mean complex Gaussian circularly symmetric with unit variance, independent and  identically distributed (i.i.d.) tone-to-tone. Figures~\ref{f:fading_scenario_sig_4QAM_diffInterfConst} and~\ref{f:fading_scenario_sig_64QAM_diffInterfConst} show the results for $M_S\!=\!4$-QAM and $64$-QAM respectively, with $M_I \!=\! \text{4-,16-,64-QAM}$ using $N\!=\!24$ resource elements. The figures show that the ML classification method has a \unit[5]{dB} gain over the nulling approach for $M_S \!=\! 4$-QAM and \unit[2]{dB} gain in the case of $M_S \!=\! 64$-QAM. Therefore, the gain of the ML classification method is largest for small constellation sizes of the desired signal, i,e. the largest gain is attained when the receiver complexity is minimal.

\begin{figure}
\centering
\includegraphics[scale=1]{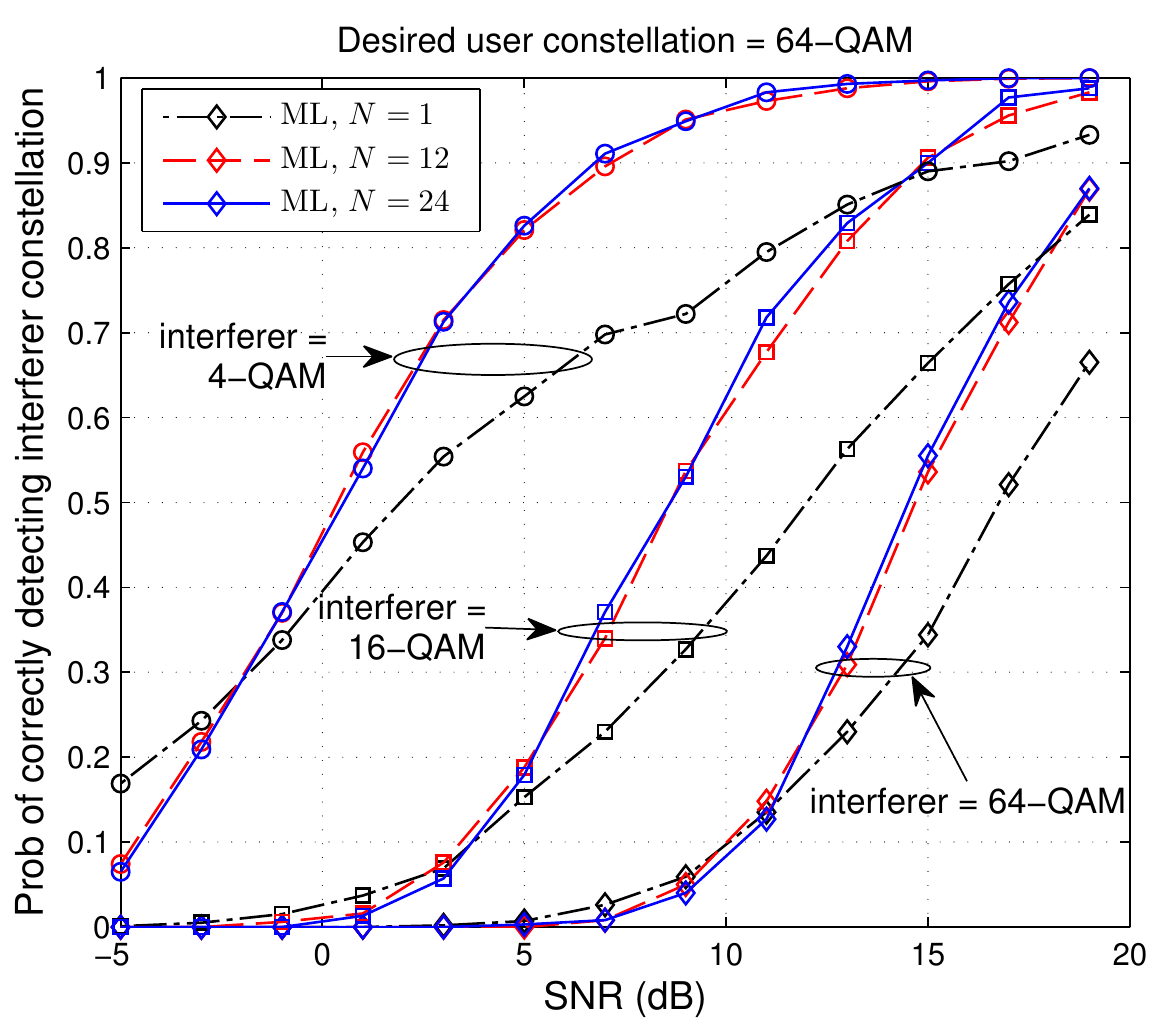}
\caption{Probability of correct interferer modulation constellation detection versus SNR (dB) for various values of $N=1,12,24$ tones and desired user is $64$-QAM. Channel is i.i.d. block fading.}
\label{f:fading_scenario_sig_16QAM_diffInterfConst_N_12vs24}
\end{figure}

Next, the performance of the ML modulation constellation classification scheme for different values of $N$ is investigated. In Fig.~\ref{f:fading_scenario_sig_16QAM_diffInterfConst_N_12vs24} we compare the constellation size detection performance with $N = 1$ (i.e. a single resource element), $12$ and $24$ for $\abs{M_S} \!=\! 4$ and different interference constellation sizes. Figure \ref{f:fading_scenario_sig_16QAM_diffInterfConst_N_12vs24} shows that $N \!=\! 12$ resource elements are adequate for robust detection of the interferer's constellation size. This will be exploited later in the receiver architecture to reduce computational complexity and memory requirements.

Remark: It could be said that increasing $N$ should always improves the performance due to better noise averaging. However, from~\eqref{eq:ML_estimate_max_log}, we find that, in addition to noise, we have residual errors in the minimization of the quadratic term in~\eqref{eq:ML_estimate_max_log}. This residual error occurs when the $\mbf{x}_{i}$ that minimizes the quadratic term is different from the originally transmitted $\mbf{x}$. The main reason for these errors is that the noise exceeds the minimum distance of the lattice constructed by the set of all symbol vectors $\mbf{x}_i$. Interestingly, we notice that the detection performance with $N = 1$ is better than that with $N = 12$ and 24 for very low SNR. This can be explained by the fact that adding more samples (increasing $N$) would add more noise than signal when the SNR is very low. Hence, it would be better if fewer samples are used for the detection.

\begin{figure}
\centering
\includegraphics[scale=1]{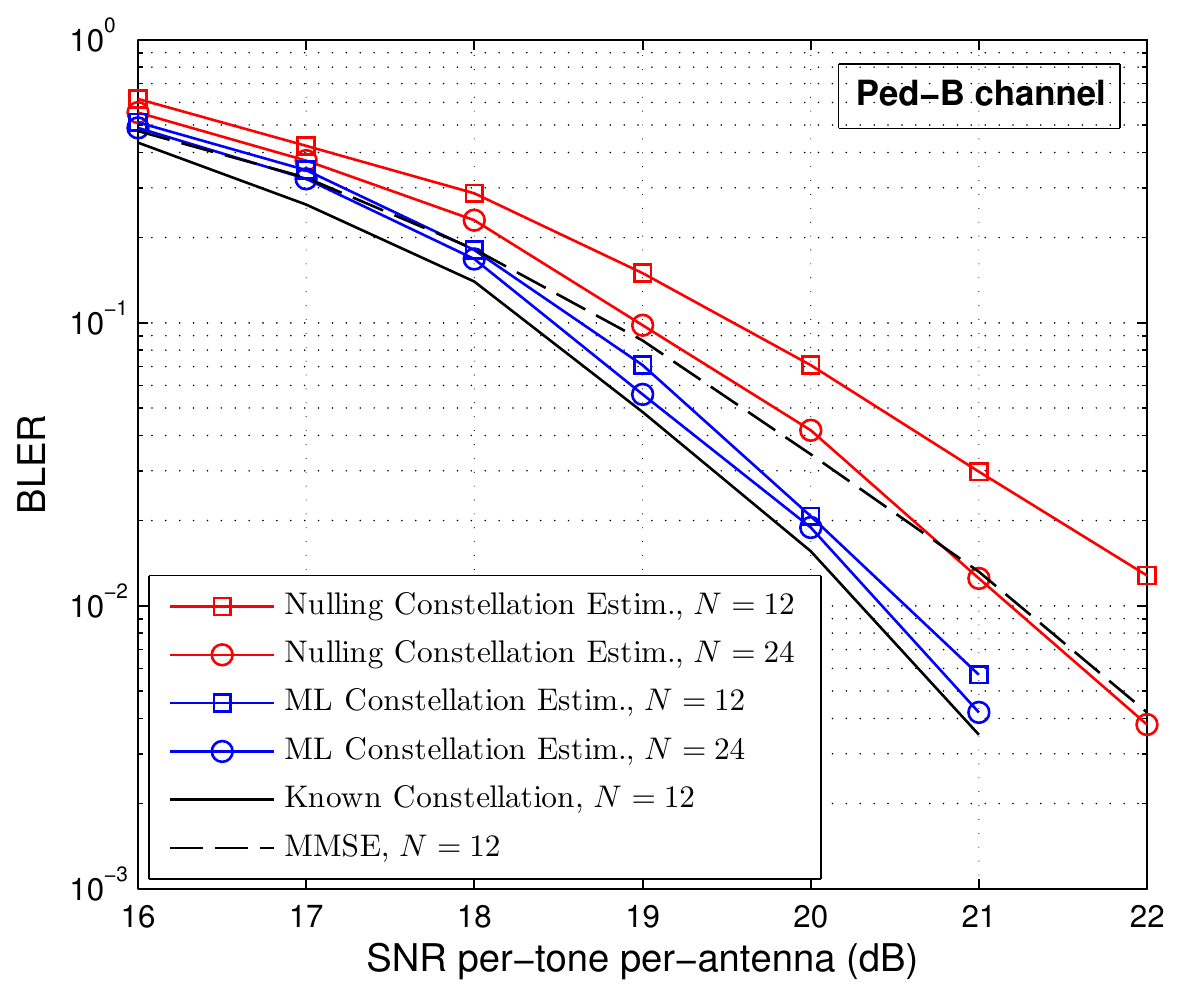}
\caption{BLER versus per-tone per-antenna SNR (dB). PED-B channel, no correlation, $64$-QAM for both users and code-rate 1/2.}
\label{f:BLER_PEDB}
\end{figure}

\begin{figure}
\centering
\includegraphics[scale=1]{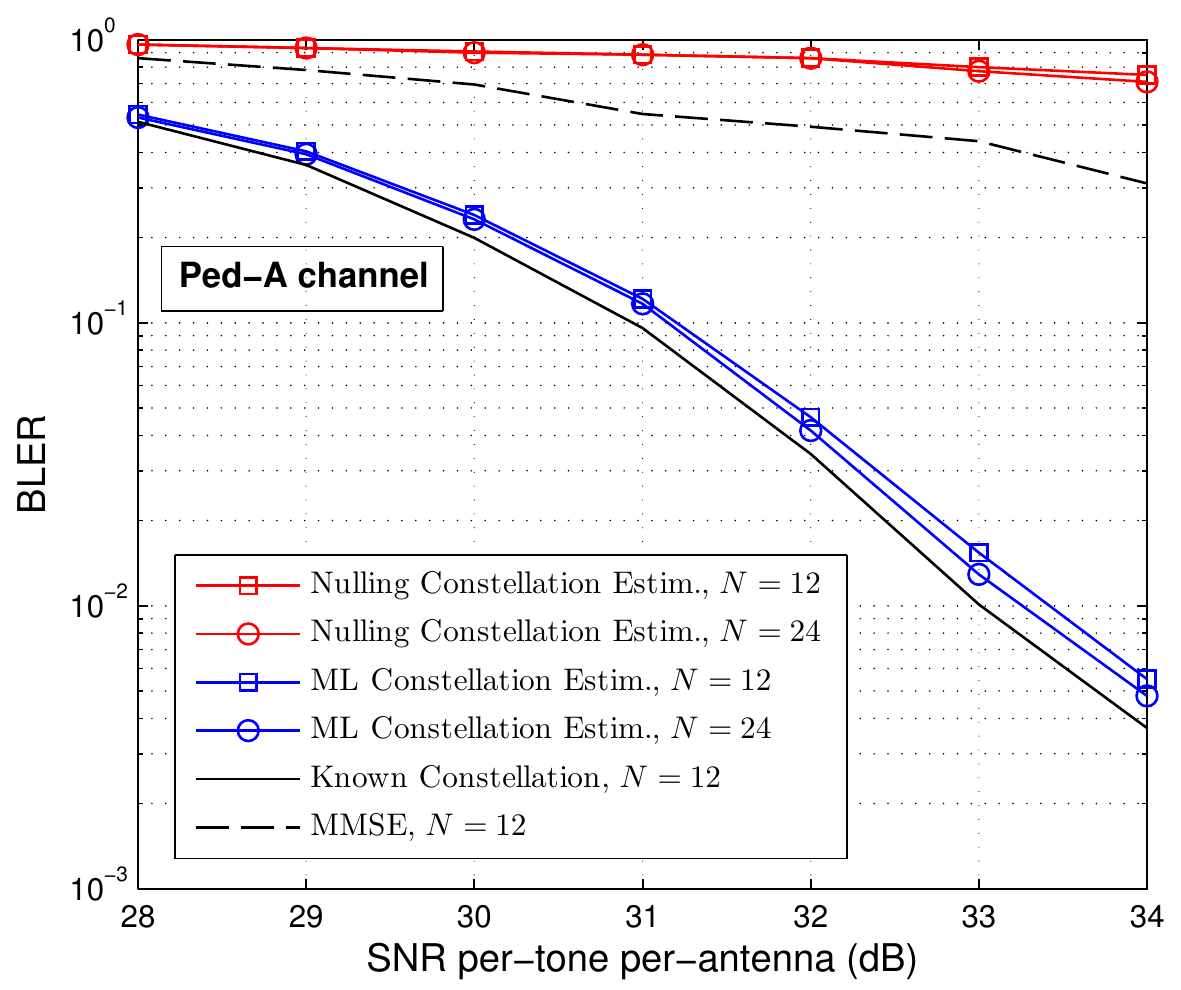}
\caption{BLER versus per-tone per-antenna SNR (dB). EPA channel, high correlation (0.9), $64$-QAM for both users and code-rate 1/2.}
\label{f:BLER_EPA}
\end{figure}

The BLER performance of the various receivers are compared when both users use $64$-QAM, with the turbo code of \cite{LTE_36.212} and encoding rate $1/2$ using block size $6144$ bits. Figure~\ref{f:BLER_PEDB} shows the results for the case of the pedestrian-B (Ped-B) multi-path channel model with no antenna correlations \cite{PedB_ChannelModel}. Figure~\ref{f:BLER_PEDB} shows that the joint ML classification and detection method is about \unit[0.1]{dB} away from having perfect knowledge of the co-scheduled user constellation. The ML method has about \unit[1.5]{dB} and \unit[1]{dB} gain over the nulling method for $N\!=\!12$ and $N\!=\!24$, respectively. The MMSE method that does not require the estimation of the co-scheduled user constellation but only its channel, has about \unit[1]{dB} degradation over the ML method. We simulated the impact of antenna correlation on the BLER performance of these receivers. The pedestrian-A (Ped-A)~\cite{PedB_ChannelModel} multi-path channel model with high antenna correlations, where both transmit and receive correlation coefficients are 0.9 is considered. The effective channel matrix is given by $\mbf{H}=\mbf{R}^{1/2}_t \mbf{H}_c \mbf{R}^{1/2}_r$, where $\mbf{H}_c$ is channel whose entries are uncorrelated and generated according to the Ped-A model, $\mbf{R}_t$ and $\mbf{R}_r$ are the transmit and receive antenna $2\!\times\!2$ correlation matrices, respectively, which have $1$ on the diagonal entries and $0.9$ on the off-diagonal. As seen in Fig. \ref{f:BLER_EPA}, the nulling and MMSE methods have significant performance degradation as compared to the joint ML method.

\section{Conclusions}\label{s:conclusion}
A receiver for MU-MIMO transmission in the context of OFDM where two users share the same time and frequency resources has been developed. This receiver is based on the joint ML modulation classification of the co-scheduled user and data detection which outperforms linear receiver approaches. It has been shown that using the max-log-MAP approximation, the decision metric for the modulation classification is an accumulation over a set of tones of the Euclidean distance metrics which is also used in the detector for LLR computations. An efficient hardware architecture emerges that exploits this commonality between the modulation classification and data detection steps and results in a sharing of hardware resources.

\newpage

\bibliographystyle{IEEEtran}
\bibliography{IEEEabrv,MIMOdetectorbib}

\vspace{0.1in}

\end{document}